\documentclass[aps,prb,twocolumn,floatfix]{revtex4-1}
\usepackage{graphicx,epsfig,array,color}
\usepackage{color,graphicx,amssymb,amsmath}
\usepackage[utf8]{inputenc}
\begin{document}
\title{Frustration effects at finite temperature in the half filled Hubbard model}
\author{Gour Jana}
\author{Anamitra Mukherjee}
\affiliation{School of Physical Sciences, National Institute of Science 
		Education and Research, HBNI, Jatni 752050, India.}
\date{\today}
	
\begin{abstract}
We investigate the finite temperature properties of the half filled Hubbard model in two dimensions, with onsite interaction ($U$), in presence of (frustrating) next nearest neighbor hopping ($t^{\prime}$) using a semiclassical approximation scheme. We show that introduction of $t^{\prime}$ results in a finite temperature pseudogapped (PG) phase that separates the small $U$ Fermi liquid and large $U$ 
Mott insulator. We map out the PG to normal metal crossover temperature scale ($T^*$) 
as a function of $U$ and $t^\prime$. We demonstrate that in the PG phase, the quadratic dependence of resistivity on temperature is violated due to thermally induced spin fluctuations. We conclude with exact diagonalization calculations, that complement our finite temperature results, and indicate the presence of a frustration driven PG state between the Fermi liquid and the Mott insulator at zero temperature as well.
\end{abstract}
	
\maketitle
\section{Introduction}
One of the fundamental questions in strongly correlated systems is the fate 
of a Fermi liquid (FL) under strong correlation and frustration effects. 
Intimately tied to this, are the understanding of important issues in materials 
theory such as, the pseudogapped phase in the doped cuprates 
\cite{dagotto-rmp-corr,plakida-review-af-pair}, non Fermi liquid (n-FL) 
behavior in the the heavy fermion compounds\cite{nfl-heavy-f,nf-heavy-f-dis} 
and rare earth nickelates 
\cite{nfl-nickelates-1,nfl-nickelates-2,nfl-nickelates-3}. With these questions 
in mind, the Hubbard model, with nearest ($t$) and next nearest hopping 
($t^{\prime}$) and its variants, have continued to be in focus of 
intense investigations, using a host of approaches such as Hartree-Fock mean 
field theory \cite{hf-ttp-0,hf-ttp-2,hf-var-ttp,hf-ttp,qmc-hf-ttp-2,VCA-ttp,VCA-ttp-2}
quantum Monte-Carlo 
\cite{qmc-ttp-1,qmc-ttp-2,qmc-ttp-3,qmc-hf-ttp-2,dqmc-ttp-1}, 
variational methods \cite{variational-wave-function-ttp}, Gutzwiller projected 
wave-function approach
\cite{Gutzwiller-projected-wvf-ttp-1,Gutzwiller-projected-wvf-ttp-2}, slave 
boson theory \cite{sl-ttp-1}, dynamical mean field theories
\cite{dmft-ttp-1,dmft1-ttp, dmft-ttp-n,dmft-ttp2, cdmft-ttp1,cdmft-ttp2} and 
effective 
spin models at large $U$ 
\cite{j1j2-rpa,j1j2-sl-dmrg,j1j2-mft-2,j1j2-cca,j1j2-pt}. 
Further, lattice implementation of time dependent density functional theory has also been used to study effect of disorder induced frustration on transport\cite{nr-1} and melting of Mott phase \cite{nr-2} in the Hubbard model.

It is well established that the Slater insulating state at half filling and weak 
interaction strength ($U$), is destabilized due to particle hole symmetry 
breaking for any non zero $t^{\prime}$  and results in a FL metal. Upon increasing $U$, this 
metal undergoes a Mott transition with either 
($\pi,\pi$) or 
($\pi,0$)/(0,$\pi$) magnetic order depending on the strength of the next nearest 
hopping $t^{\prime}$. The investigation of $t^\prime$ induced metallic state 
at zero temperature using dynamical cluster approximation (DCA) in the paramagnetic phase 
\cite{dmft1-ttp, 
dmft-ttp2} and cluster-dynamical mean field theory (CDMFT) 
\cite{cdmft-ttp1,cdmft-ttp2} have lead to the following major conclusions.
For small $t^{\prime}/t\sim -0.3$, increasing $U$ causes the metal to undergo 
a 
two stage transition. Instead of directly going from a metal to a Mott insulator, 
there is an intermediate regime where the Fermi surface is gapped out along 
the 
($\pi,0$) and ($0,\pi$) directions, while the ($\pi,\pi$) direction remains 
gapless. The total density of states (DOS) shows a pseudogap and the metal 
is 
predicted to be a n-FL. Further, renormalization group studies 
\cite{rg-ttp,pom-5-van,pom-6} suggest possible 
tetragonal or $C_4$ rotation symmetry breaking of the Fermi surface, as has 
also been found in the studies of the $t-J$ model \cite{pom-tj1,pom-tj2,pom-tj3}. 
In spite of these important advances, there are a very few results on finite 
temperature properties\cite{tremblay-3} and nature of the metallic state  
stabilized by frustration\cite{dmft1-ttp,dmft-ttp2}. Both (CDMFT)\cite{tremblay-3} and (DCA)\cite{dmft1-ttp,dmft-ttp2} assume a paramagnetic background and suffer from well known analytic continuation issues (as they are typically formulated in imaginary time). Thus, the role of magnetic background and impact of temperature has remains inadequately understood. 

Solving the Hubbard model without particle-hole symmetry using standard DQMC approach is limited to high temperatures due to fermion sign problem. Further in DQMC sampling over the imaginary time and spatial coordinates with good accuracy is numerically extremely prohibitive. In addition analytic continuation issues for extracting real frequency quantities like density of states are well known. 
Given the rather challenging nature of the problem, in the present paper we employ a recently developed semiclassical Monte-Carlo (s-MC) that is free of fermion sign problem and is computationally inexpensive, allowing access to large system sizes. Being semi-classical in nature, the approximation agrees well with DQMC in the thermally dominated phase and has been studied and compared against DQMC in our previous publication \cite{mf-mc,ptca} for the square lattice Hubbard model at half filling. Given this background, it is natural to test the method to extract finite temperature properties of the rather difficult $t-t^\prime$ Hubbard model in two dimensions. Also, while previous studies have heavily focussed to $t^\prime/t\sim -0.3$, a value relevant for the cuprates, here we study the effect of frustration for the entire window of $|t^\prime/t|\in [0,1]$. 

We begin by summarizing our main results. We first discuss the impact of frustration on the magnetic phases at finite temperatures. 
This includes, suppression of the $t^{\prime}=0$, G-type $\mathbf{q}=(\pi,\pi)$ magnetic transition temperature, shifting of the regime of preformed local moments to larger $U$ values and emergence of A-type ($\mathbf{q}=(\pi,0)$ or $\mathbf{q}=(0,\pi)$) order at large frustration. We then present the $T-t^\prime$ metal insulator phase diagrams at different $U$ values to establish the existence of finite temperature PG metal and determine the temperature scale ($T^{*}$) for the PG to normal metal crossover. From the dependence of resistivity on temperature, we show a continuous crossover from a small $U$ Fermi liquid (FL) to the PG phase (with the resistivity temperature exponent deviating strongly from 2). We show how the interplay of frustration and correlation drives local moment fluctuation at finite temperature and stabilizes the PG metal in the passage from the FL to a Mott state.  We close by providing $T=0$ ED results, that establishes the same phenomenology of a PG state punctuating the FL to Mott transition with increasing interaction strength, in presence of frustration in agreement with DCA results\cite{dmft1-ttp, 
dmft-ttp2}. 

The paper is organized as follows. In Sec. II we briefly discuss effective Hamiltonian derived from the many body problem. In Sect. III we present our main results and conclude the paper in Sec. IV.

\section{Model $\&$ method}
The $t-t^{\prime}$ Hubbard model has the following form:
\begin{eqnarray}
H&=&H_o+H_1\\\nonumber
&=&-t\sum_{\langle i,j \rangle,\sigma}(c^{\dagger}_{i,\sigma} 
c^{\phantom{\dagger}}_{j,\sigma} + h.c.) \\\nonumber
&-&t^{\prime}\sum_{\langle\langle i,j 
\rangle\rangle,\sigma}(c^{\dagger}_{i,\sigma} 
c^{\phantom{\dagger}}_{j,\sigma} + 
h.c.)+U\sum_i n_{i,\uparrow}n_{i,\downarrow} 
\label{1}
\end{eqnarray}
Here the model is defined on the two dimensional square lattice with $t$ being the nearest neighbour hopping and $t^{\prime}$ being the next nearest neighbor hopping. $U$ is the correlation strength and $H_o$ and $H_1$ denote the kinetic and interaction Hamiltonians respectively. 

The (s-MC) approach\cite{mf-mc,ptca}, is based on Hubbard-Stratonovich (HS) decomposition of the interaction Hamiltonian by introducing auxiliary fields (Aux. F.) just as in Determinantal Quantum Monte Carlo (DQMC). To apply the Hubbard Stratonovich (HS) decomposition, we first write the local 
interaction term  as a sum of squares of total on-site density and spin 
operator. In the Appendix subsection A, we provide detailed derivation of the 
effective one body Hamiltonian ($H_{eff}$), that is obtained from the many body problem 
after employing HS transformation and retaining only the temperature induced 
spatial fluctuations of the (Aux. F.)s. 
For continuity of the main paper, we mention that two (Aux. F.) are introduced, a vector field $\mathbf{m}_i$ and a scalar field $\phi_i$ at every site of the lattice $i$. They couple to the spin and the charge degrees of freedom respectively. With the introduction of these fields, as shown in the Appendix, we get the following effective Hamiltonian, which is used in the paper:

\begin{eqnarray} 
H_{eff}&=&
H_o+\frac{U}{2} \sum_i(\langle n_i\rangle n_i-{\bold m_i}.{\bold \sigma_i})
\label{ham}\\
&+&\frac{U}{4}\sum_i({\bold m_i}^2-\langle n_i \rangle^2)-\mu\sum_i n_i
\nonumber
\end{eqnarray}

Here, $\sigma_i$ denote the three Pauli matrices and are related to the spin operator at a site $i$ by $\mathbf{S_i}=\frac{1}{2}\sigma_i$, as discussed in Appendix A.
We confine our calculations to the regime of  $t^{\prime}/t<0$, to compare with cuprate inspired literature that have chosen $t^\prime/t<0$. We will measure all energy scales in units of the nearest neighbour hopping $t$. We note that the effective Hamiltonian looks identical to a simple (mean field) Hartree Fock model. The crucial difference of (s-MC) from a finite $T$ mean field theory is that, the Aux. F. appearing in $H_{eff}$ are treated within a classical Monte Carlo instead of using self consistency or saddle point approach, as would be done in a mean field treatment. The thermal sampling brings in the effects of spatially inhomogeneous thermal fluctuations. These thermal fluctuations capture many of the well established features such as the regime of preformed local moments, non-monotonic dependence of $T_N$ on $U$ as found in DQMC in the unfrustrated problem\cite{mf-mc}. These results cannot be produced in an unrestricted finite $T$ Hartree-Fock mean field theory. However, at low temperatures, when quantum fluctuations dominate, the s-MC reduces to a uniform Aux. F. solution identical to Hartree-Fock mean field theory as the temporal (quantum) fluctuations of the Aux. F. is not taken into account. This limits the validity of the s-MC only to thermally dominated regime. 

The calculation scheme for s-MC is the following. At any fixed temperature, the goal of the s-MC is to generate equilibrium configurations of Aux. F. through a Monte Carlo scheme. The coupled one body quantum problem is exactly diagonalized for a fixed Aux. F. background and the total free energy is used to update the Aux. F. within a Metropolis algorithm. The method of classical Monte-Carlo coupled with exact diagonalization used to solve Eq. 2 is  detailed in Appendix subsection B. Here we, very briefly, outline some relevant details. We start by choosing a background configuration of (Aux. F.) at a high temperature, then diagonalize the system in that (Aux. F.) background. We then update the (Aux. 
F.), 
re-diagonalize the coupled fermion problem. The free energy difference is used 
to 
accept the (Aux. F.) update. After adequate thermalization system sweeps 
(involving update attempts by visiting each site sequentially), we generate 
equilibrium configurations of (Aux. F.) from which observables are calculated. 
As in classical Monte-Carlo, we typically anneal down from high to low temperatures.
A number of indicators were calculated, density of states (DOS) 
$N(\omega)$, static spin structure factor $S_q$, optical conductivity $\sigma(\omega)$ and quantum local moment distribution $P(M)$. The definitions of these standard indicators are given in the Appendix subsection C. Computational codes for (s-MC) were developed in house.

\section{Results}
%

\textit{1. \underline{Evolution of magnetic phases with temperature:}}

\begin{figure}[t]
\centering{
\includegraphics[width=8.5cm, height=8.25cm, clip=true]{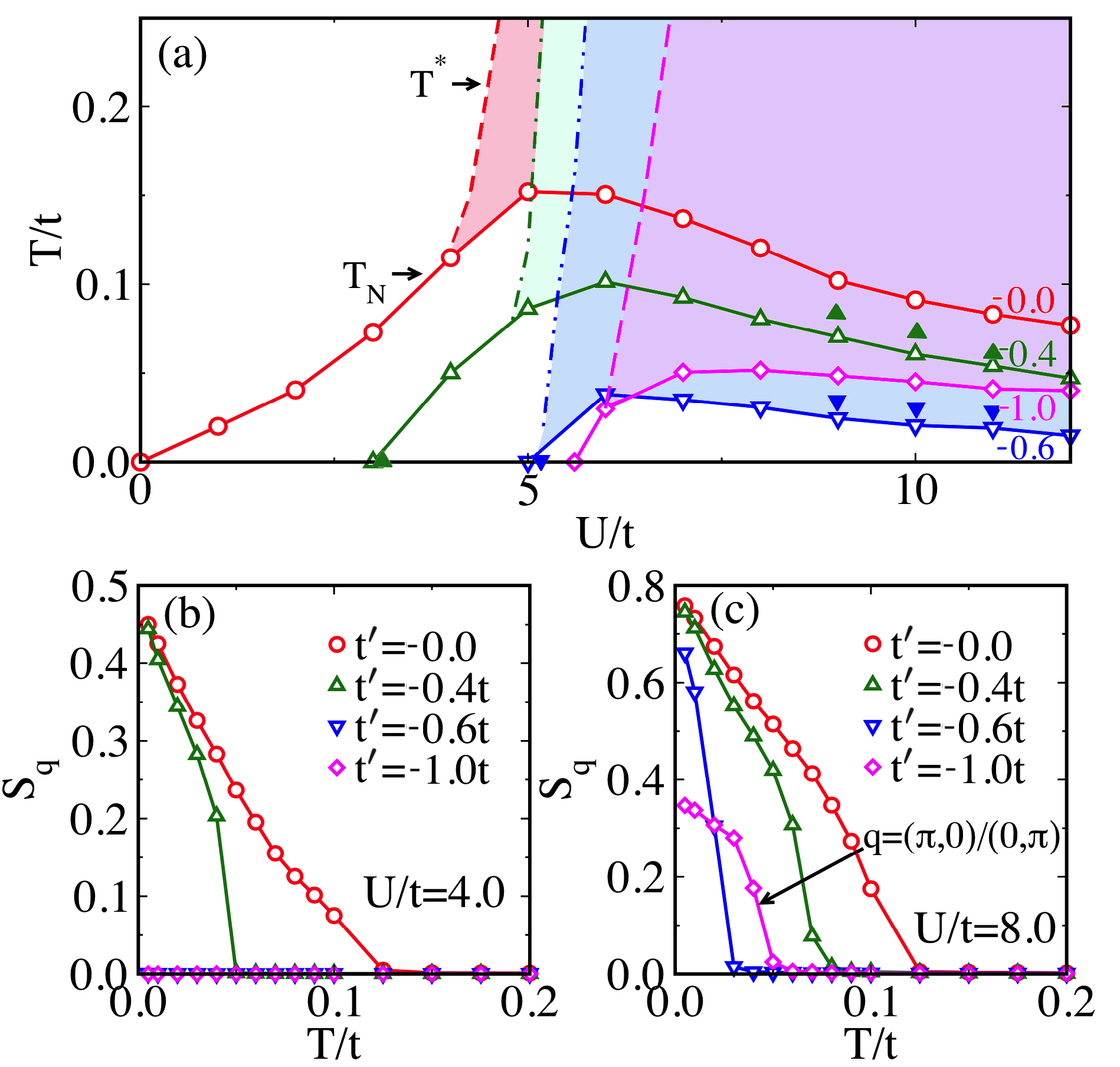}}
\caption{Color online: (a) shows the $T_N$ for various $t^{\prime}$ values as function of $U/t$. The $t^{\prime}$ values are marked on the curves. For $t^{\prime}/t=0$, -0.4 and -0.6, the magnetic phase is G-type, while for $t^{\prime}/t=-1$, the magnetic order is A-type (the magenta with diamonds). The solid symbols represent corresponding magnetic transitions from three dimensional results on 8$^3$ systems for $t^\prime/t=-0.4$ (solid up triangles) and $t^\prime/t=-0.6$ (solid down triangles) at large $U$. The regions to the right the various dashed and dot-dashed lines are preformed local moments regimes. The dashed and dot-dashed line and the shading of region to the right are color coded to the corresponding $T_N$ shown in open symbols. (b) and (c) show the static magnetic structure factors for $U/t=4$ and $U/t=8$ respectively, for different $t^{\prime}$ values as indicated. In (b) we show the \textbf{q}$=(\pi,\pi)$ case only while in (c) the magenta line (with diamonds) show the \textbf{q}$=(0,\pi)/(\pi,0)$ structure factor, rest of the curves are for \textbf{q}$=(\pi,\pi)$.}
\vspace{-0.0cm}
\label{f-1}
\end{figure}

We discuss the  $U-T$ phase diagram, showing the magnetic phases for 
different $t^{\prime}$ values in Fig.~\ref{f-1} (a). Before proceeding we make the following important cautionary note. The Mermin-Wagner theorem states that in 1D or 2D systems with SU(2) invariant Hamiltonian with short range spin interactions, magnetic ordering scale tends to zero at any finite temperature in the thermodynamic limit. Given this, the magnetic transition temperatures quoted in the present paper should be understood as crossover temperature scales at which the magnetic spin-spin correlation length extends over the system size.

The $T_N$ for \textbf{q}$=(\pi,\pi)$ or G type antiferromagnetic order at $t^{\prime}/t=0$, is shown by the solid red line with circles in panel (a). For $t^{\prime}/t=0$, we see 
the expected non monotonic behavior of $T_N$ with $U/t$.  The dashed (red) line demarcates the crossover between the preformed local moment regime with pseudogapped DOS for $T_N<T<T^*$ and paramagnetic metal for $T>T^*$. The preformed local moment extends over the entire region between the red dashed line and the $T_N$ curve. The $t^{\prime}=0$ case, studied by us previously, shows that the s-MC approach goes much beyond simple finite temperature Hartree-Fock calculations and can capture the $t^2/U$ scaling of $T_N$ at large $U/t$ and the existence of the preformed local moment  state above $T_N$. For detailed results on temperature dependence of specific heat, local moments, double occupation as well as solution of the model on very large lattices (256$^2$) in 2D and ($40^3$) in 3D, we refer to our earlier work\cite{mf-mc, ptca}.

With increasing magnitude of $t'$, the frustration increases. We find that this causes an an overall suppression of  $T_N$ for G type magnetic order and the shifts the regime of preformed local moment (broken line of same color) to higher $U$ values. 
We find that within numerical accuracy, the G type magnetic order is suppressed to zero, for $t^\prime/t\sim-0.8$. 
Beyond $t^\prime/t=-0.8$, there is an emergence of A type antiferromagnetic order. Typical $T_N$ data for A type antiferromagnetic order (magenta diamonds) and the corresponding $T^*$ (magenta dashed line) are shown for $t^{\prime}/t=-1$ in panel (a). 
We also show $T_N$ data for three dimensional 8$^3$ systems with solid symbols for $t^\prime/t=-0.4$ (solid up triangles) and $t^\prime/t=-0.6$ (solid down triangles) at large $U$, to show the qualitative correctness of 2D calculations. 

Panels (b) and (c) show supporting magnetic structure factor data. The magnetic structure factors for $U=4t$ in Fig.~\ref{f-1} (b) shows the gradual suppression of the \textbf{q}$=(\pi,\pi)$ magnetic order with increasing $t^{\prime}$, eventually leading to a paramagnet (PM). In panel (c), we see that at $U=8t$ the suppression in the \textbf{q}$=(\pi,\pi)$ order and an eventual formation of \textbf{q}$=(\pi,0)$ or $(0,\pi)$ order with increasing magnitude of $t^{\prime}$. The regime of preformed local moments for $T_N<T<T^*$, at finite $t^{\prime}$, is discussed next.

\textit{2. \underline{Finite $T$ phases in presence of frustration:}} 

\begin{figure}[t]
\centering{
\includegraphics[width=8.5cm, height=3.5cm, clip=true]{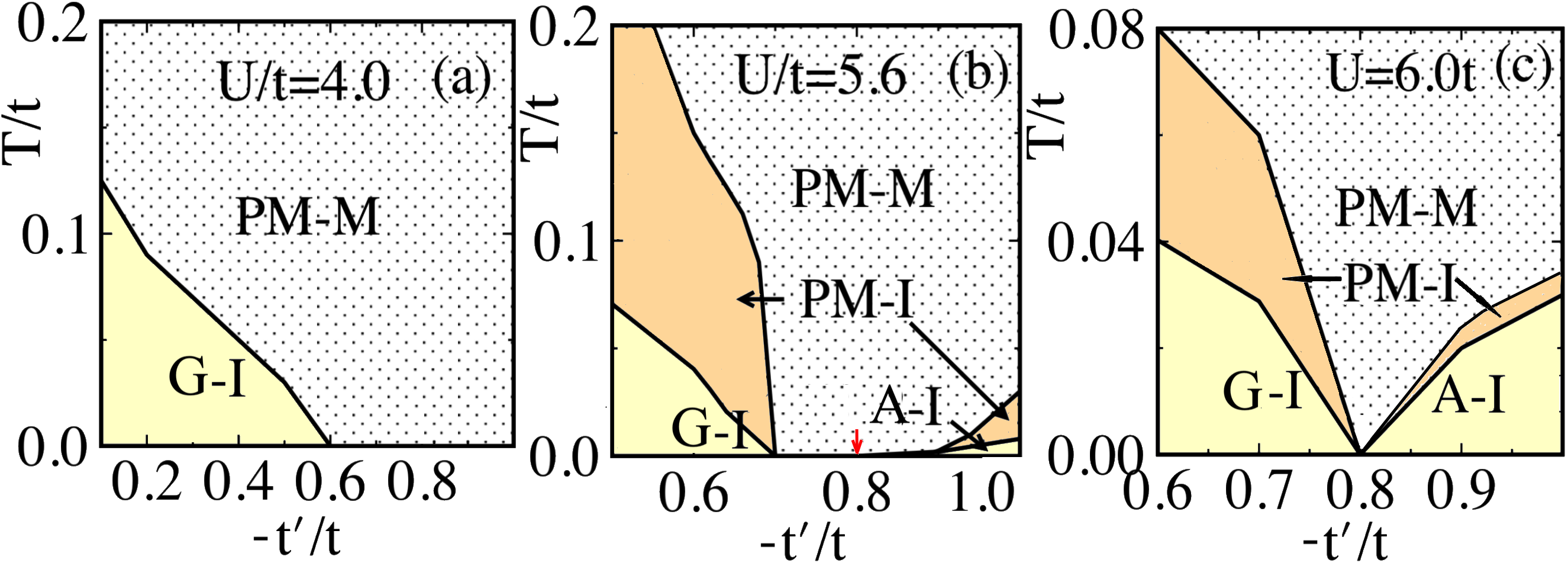}}
\caption{Color online: The temperature vs $t^{\prime}$ phase diagrams for different $U$ values as indicated in panels (a) to (c). G-I, A-I, PM-M and PM-I are respectively, G-type insulator, A-type insulator, paramagnetic metal and paramagnetic insulator. The entire PM-M is a pseudogapped metal for panels (b) and (c). The PG in the PM-M is limited to small $t^\prime$ values in (a). The finite temperature boundaries are discussed in the main text.}
\vspace{-0.0cm}
\label{f-2}
\end{figure}

To extract the impact of frustration of the thermal evolution of $t^\prime=0$ insulating states, in Fig~\ref{f-2} (a) to (c) to show the $T-t^{\prime}$ phase diagrams at three $U$ values\cite{low-temp}. The plots are shown for increasing values of $U$. While for $t^\prime=0$, the ground state of all the three cases are insulating antiferromagnets, we find strikingly different thermal response in presence of frustration. As discussed below, these phase diagram bring out the contrasting effect of frustration on weak and strong correlation situations. 

The finite $T$ scale for metallization of the insulating states are determined by studying the behavior of the optical conductivity ($\sigma(\omega)$) with $\omega$. For a metal there has to be a constant DOS close to the Fermi energy. Thus, $\omega\sigma(\omega)$ should have a linear dependence on $\omega$ as 
$\omega\rightarrow 0$. For (charge gapped) insulator there is typically a gap in $\sigma(\omega)$ up to some finite frequency starting from $\omega=0$. The details of the optical conductivity calculations are discussed in the Appendix subsection D. Here we focus on the main results.

We find that at smaller $U(=4t)$ values, the temperature scale for  metallisation of the G type Mott insulator is monotonically suppressed with increasing $t^{\prime}$. The G type magnetic order is also lost simultaneously with loss of the insulating nature. The non magnetic metal remains stable all the way up to the maximum possible value of frustration ($t^{\prime}/t=-1$) and is devoid of local moments except for very small frustration values. In sharp contrast, on increasing $U$, in panel (b), correlation effects conspire with frustration to form a new Mott state with A type magnetic order, which is most stable for $t^{\prime}/t=1$ (with the largest metallisation temperature scale as seen in panel (b)). This A type insulator invades the metallic regime, essentially limiting the metal to finite $t^{\prime}$ regime at low $T$ and fanning out with increasing temperature. With further increase in $U$, in panel (c), the metallic regime shrinks to a point at $t^\prime/t=-0.8$ (within numerical accuracy). 

With temperature increase, for the larger $U$ cases in panels (b) and (c), the antiferromagnetic insulator first gives way to a PM insulator and then to a PM-M. This is because at larger $U$, local moments form at a temperature higher that the magnetic ordering scale. Although we do not report it here, we find a two peak specific heat structure one corresponding to the moment formation and the other to moment ordering temperature, similar to the unfrustrated case\cite{mf-mc}. The effect of local moments on the nature and transport properties of the metal at finite temperature is discussed next.


\begin{figure}[t]
\centering{
\includegraphics[width=8.5cm, height=8.5cm, clip=true]{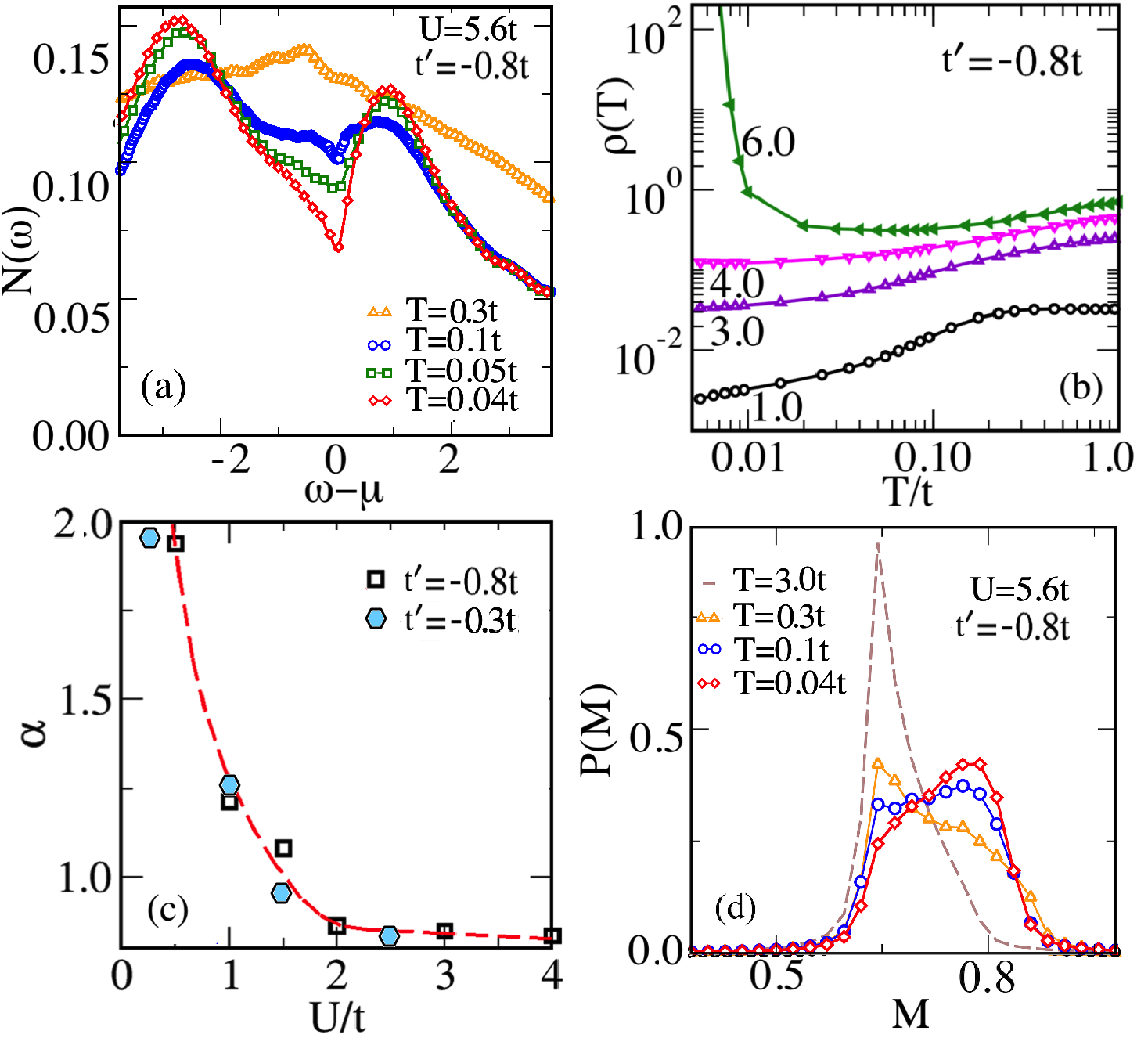}}
\caption{Color online: Panel (a) presents the evolution of DOS with temperature and shows the pseudogapped to non-pseudogapped crossover for $t^\prime/t=-0.8$ and $U/t=5.6$. (b) shows the resistivity, $\rho(T)$ measures in units of $\pi e^2/\hbar a$, for different $U$ values for $t^{\prime}/t=-0.8$. (c) shows the plot of the temperature exponent $\alpha$ from fitting $\rho=AT^{\alpha}+B$ as a function of $U/t$ for the metallic cases for $t^{\prime}/t=-0.8$ (open squares) and for $t^\prime/t=-0.3$ (filled hexagons). (d) shows the distribution of magnetic moments $P(M)$ as a function of temperatures corresponding data in panel (a).}
\vspace{-0.0cm}
\label{f-3}
\end{figure}

\textit{3. \underline{Pseudogapped metal at finite $T$:}} 

We now we focus on the behavior of the frustration induced metal at finite temperature. We would like to emphasize that the (s-MC) method can provide reliable results in thermally dominated regime. At low temperatures, where quantum fluctuations are important, it reduces to inhomogeneous Hartree-Fock. Thus in this section we provide the s-MC results and discuss $T=0$ results from exact diagonalization in the next section.  
For demonstration, we show data at $t^\prime/t=-0.8$ and $U/t=5.6$. This point is marked in by the small red arrow in Fig.~\ref{f-2} (b).

Fig.~\ref{f-3} (a) shows the DOS at different temperatures for this metallic state. We see that, the non-pseudogapped DOS at very high temperature (orange curve with triangles) develops a PG with temperature decrease. Further, the PG deepens with reducing temperature. For this $U$ value of $T_N/t\sim0.04$, thus the DOS are shown for $T\geq T_N$, where thermal fluctuations dominate. Such pseudogap feature is seen for the entire metallic regime at $U/t=5.6$ and $U/t=6$ in Fig.~\ref{f-2} (b) and (c) respectively. For $U/t=4$, the PG is restricted to small $t^\prime$ values. Panel (b) in Fig.~\ref{f-3} shows the resistivity as a function of temperature at $t^{\prime}=-0.8t$ and different $U$ values. We see a transition from an insulator for $U/t\geq 6$ to a metal for $U/t<6$. The $T^*$ scales in Fig.~\ref{f-1} (a) are defined to be the highest temperature $T$($>T_N$) where the DOS develops a local minima at $\omega-\mu=0$. 


Panel (c) shows the exponent $\alpha$ of the temperature dependence of $\rho(T)\equiv AT^{\alpha}+B$, obtained from fitting resistivity for the metallic cases\cite{range} for $t^{\prime}/t=-0.8$ (open squares) and for $t^\prime/t=-0.3$ (filled hexagons). We see that $\alpha=2$, only for small $U$ values. Beyond $U/t=0.5$, there is a sharp drop in the value of the exponent, and it saturates to a sub linear value. $\alpha$ continues to be remain fixed at the sub linear value till the Mott state is reached at $U/t=5.6$ for $t^\prime/t=-0.8$ and at $U/t=3$ for $t^\prime/t=-0.3$.
\begin{figure}[t]
	\centering{
		\includegraphics[width=8.5cm, height=4.5cm, clip=true]{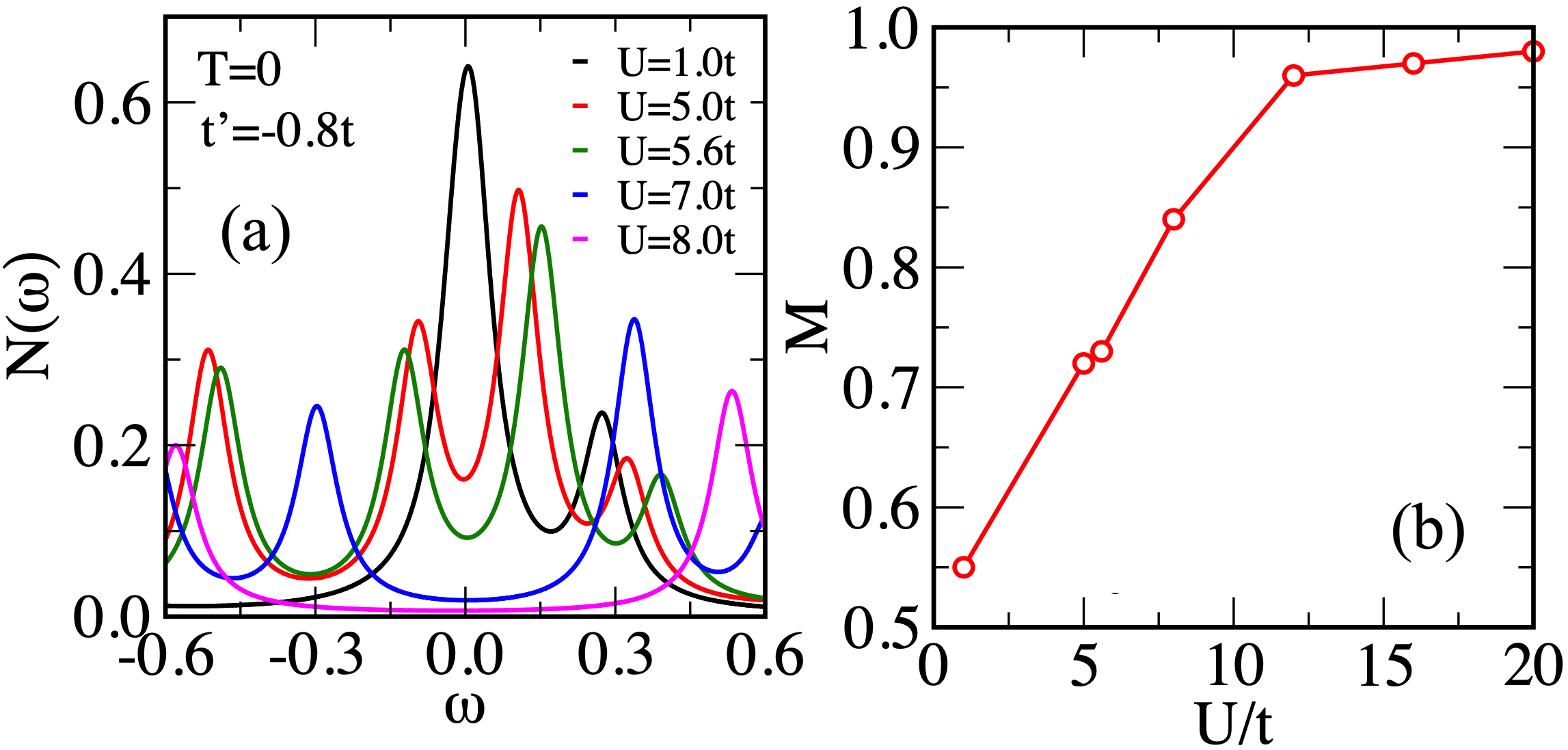}} 
	\caption{Color online: (a) Single particle DOS  at $T=0$ from Lanczos on $4\times 6$ system. The data is shown for $t^\prime=-0.8t$, for different $U$ values. (b) shows the local  moment ($M$) as a function of $U/t$ at $T=0$ for $t^{\prime}=-0.8t$. These results were generated using Lanczos code developed in house.}
	\vspace{-0.0cm}
	\label{f-4}
\end{figure}

To investigate the cause of the deviation of $\alpha$ from 2, in panel (d) we show the real space distribution of local spin moments, $P(M)$ with decreasing $T$. The data is shown at $t^\prime/t=-0.8$ and $U/t=5.6$ that is close to the Mott insulator at $U=6t$. Here $M=\langle(n_{\uparrow}-n_{\downarrow})^2\rangle$, which is equivalent to $1-2\langle n_{\uparrow}n_{\downarrow}\rangle$ at half filling. Thus, the value $M=0.5$ implies no local moments, as $\langle n_{\uparrow}n_{\downarrow}\rangle=0.25$ for $U=0$. For the high temperature case, the DOS is non-pseudogapped and the $P(M)$ has a dominant peak at around 0.5 (brown dashed line) indicating almost no local moments. Where as in the pseudogapped regime, the $P(M)$ is highly non uniform.  This indicates that once local moments start forming in the vicinity of the Mott state, frustration makes their spatial distribution non uniform. The ensuing scattering of the fermions from these spatially fluctuating moments causes the deviation from the FL behavior for the resistivity. As temperature is lowered further, we see that the moment distribution in real space begins to peak at about $0.8$. As $T\rightarrow 0$, this peak is expected to grow and become uniform, which would be the Hartree-Fock result. In this limit the s-MC method will show a non-pseudogapped state. Thus, as temperature is lowered if the PG is to survive the theory needs to incorporate quantum fluctuations. For this we turn to the $T=0$ limit in the next section. We also note that similar deviation of $\alpha=2$ and pseudogap occurs near the metal to Mott insulator boundary for all $t^\prime/t\in[-0.2,-1]$. Values of $\alpha$ for small $t^\prime=-0.3$ are shown by filled hexagons in Fig.~\ref{f-3} (c). For this case the Mott state occurs beyond $U/t=2.5$. For $-0.2<t^\prime<0$, it is difficult to numerically claim a PG phase.

\textit{4. \underline{Pseudogapped metal at $T=0$:}} 

So far we have established the existence of a finite $T$ PG metal and its origin has been shown to be spatial fluctuations of local moment. As mentioned above, it is expected that at low $T$, quantum fluctuations will dominate. In this regime s-MC produces uniform moment static solutions, which can not host a pseudogapped state. In Fig.~\ref{f-4} we present $T=0$ Lanczos based ED density of states in panel (a) and evolution of local moment size in panel (b) on small ($4\times 6$) clusters. The data is shown for $t^{\prime}/t=-0.8$ and different $U$ values. We see that the (normal) non-PG DOS at small $U$, develops a pseudogap for $U/t=5$ to $7$. In this $U$ range, the local moment size grows from 0.5 (uncorrelated value) to $\sim 0.8$. This shows a clear correlation between existence intermediate sized local moment and the pseudogapped state at $T=0$. The PG eventually hardens in to a Mott gap once the moment size grows beyond $0.8$. This $T=0$ quantum fluctuation driven PG crosses over to the s-MC thermal fluctuation induced PG. This large $t'/t$ result complements previous $T=0$ results at small $t'/t(=-0.3)$ values\cite{dmft1-ttp,dmft-ttp2}.

\section{Conclusions $\&$ Discussion}

In this paper we have used a semiclassical-Monte-Carlo approach to study the 
effects of correlation and frustration in the two dimensional  Hubbard model at half filling at finite temperature. We have studied finite $T$ evolution of magnetic phases, metal insulator transitions and have mapped out the temperature scale ($T^*$) of the crossover of the PG regime to normal metal. Our finite temperature s-MC and $T=0$ Lanczos results are consistent and show that the Fermi liquid to Mott 
transition with increasing $U$, in presence of intermediate to large 
frustration ($t^\prime$), is not a direct transition,  
but goes through a pseudogapped metallic phase phase. While we have focussed at $t^\prime/t=-0.8$, we have found numerically that this phenomenology holds for a $t^\prime/t$ between -0.2 to -1. 

As mentioned in the paper, the goal of the s-MC method is to generate equilibrium configurations of Aux. F. through a classical Monte Carlo at given temperature. 
The acceptance of an attempted update of Aux. F. depends on both the existing classical Aux. F. background as well as the quantum mechanical fermion problem. 
Away from the metal insulator boundary, we find uniform Aux. F. configurations minimize the free energy leading to a small $U$ uniform Fermi liquid metal or a large $U$ Mott state with spatially uniform local moments. At finite temperature, and close to the phase transition, the Aux. F.s access free energy minima corresponding to the Fermi liquid and the Mott insulator due to thermal fluctuations and stabilize inhomogeneous local moments which scatter against fermions leading to the finite temperature pseudogapped state. At very high temperature ($T>T^*$), the Aux. F. decouple from the fermions, as was demonstrated in our earlier work on the unfrustrated Hubbard model\cite{mf-mc}, and the pseudogapped state disappears. The method works well in this intermediate to high temperature regime where thermal fluctuations dominate the physics. At sufficiently low temperature as it misses out on quantum fluctuations and reduces to inhomogeneous Hartree-Fock limit. The determination of the thermal to quantum fluctuation dominated crossover as a function of temperature is, at present, an open question. Due to this our strategy is to provide complementary $T=0$ Lanczos results that, within size limitations, indicate frustration driven pseudogapped state between the Fermi liquid and the Mott insulator.

In conclusion, these results contribute to the long standing questions 
of the nature of strong interaction driven metal-insulator transitions in the 
background of frustration and are of relevance to many body theory and 
materials physics alike.

.

\vskip -0.2cm
\section{Acknowledgement} 
We acknowledge the KALINGA, NOETHER and XANADU computational clusters at NISER. We also acknowledge, Nitin Kaushal for helping set up the ED code.

\vskip 2cm
\begin{center}
\textbf{APPENDIX}
\end{center}
In the subsection A we discuss the derivation of the $H_{eff}$ used in the 
main paper. In subsection B we present the technical details of solution 
methodology and in subsection C, we define the various indicators used to 
study the effective Hamiltonian. 
\subsection{Derivation of $H_{eff}$}
\begin{eqnarray} 
n_{i,\uparrow}n_{i,\downarrow} =\frac{1}{4}(n_i^2)-({\bold S}_i \cdot 
\hat{\Omega}_i)^2.
\label{2}
\end{eqnarray}
Here, the spin operator is ${\bold S_i}=\frac{\hbar}{2}\sum_{\alpha,\beta} 
c^{\dagger}_{i,\alpha}
{\bold 
\sigma}^{\phantom\dagger}_{\alpha,\beta}c^{\phantom\dagger}_{i,\beta}$, 
$\hbar=1$,  
$\{\sigma^x,\sigma^y,\sigma^z \}$ are the Pauli matrices, and $\hat{\Omega}$ 
is 
an 
arbitrary unit vector. In the previous identity, we have used 
the fact that 
$({\bold S}_i\cdot \hat{\Omega}_i)^2=(S_{i,x})^2=(S_{i,y})^2=(S_{i,z})^2$.  
This rotation invariant decoupling results in the  correct Hartree-Fock saddle 
point 
after implementing a Hubbard-Stratonovich (HS) decomposition. We start with 
the 
partition function $Z=Tre^{-\beta H}$ where the trace is over all particle 
numbers 
and site occupations. $\beta=1/T$, with $k_B$ set to 1. We divide the interval 
$[0,\beta]$ into $M$ equally spaced slices, defined by $\beta=M\Delta \tau$, 
separated by $\Delta \tau$ and labeled from 1 to $M$. For large M, we employ 
the usual Suzuki-Trotter decomposition, to write 
$e^{-\beta(H_o+H_1)}=(e^{-\Delta 
\tau H_o}e^{-\Delta \tau H_1})^M$ to first order in $\Delta \tau$. 
From Eq. (2) and the HS identity, $e^{-\Delta \tau U \sum_i 
[\frac{1}{4}(n_i^2)-({\bold S}_i \cdot \hat{\Omega}_i)^2]}$, for any time slice 
$^\prime l^\prime$, is found to be proportional to,
\begin{eqnarray} 
\int {d\phi_i(l) d\Delta_i(l)
	d^2\Omega_i(l)}\times \hskip 4.4cm& \nonumber\\
e^{-\Delta \tau [\sum_i(\frac{\phi_i(l)^2}{U}+i\phi_i(l)n_i+\frac{{\Delta_i(l)}^2}{U}
	-2 {\Delta_i(l)}\hat{\Omega}_i(l).{\bold S_i})]}&\nonumber
\end{eqnarray}
Here two auxiliary fields, $\phi_i(l)$ that couples to the local charge density, and 
$\Delta_i(l)$ that couples to the spin density are introduced. Defining the 
product 
$\Delta_i(l)\hat{\Omega}_i(l)$ as a new vector auxiliary field, ${\bold m_i}(l)$ at 
every site we can write the partition function as:
\begin{widetext}
	\begin{equation}
	Z= const. \times Tr \prod^1_{l=M} \int {d\phi_i(l) d^3m_i(l)}e^{-\Delta \tau 
	[H_o+\sum_i(\frac{\phi_i(l)^2}{U}+i\phi_i(l)n_i+\frac{{\bold m_i(l)}^2}{U}
		-2 {\bold m_i(l)}.{\bold S_i})]}
	\end{equation}
\end{widetext}
The integrals are over the auxiliary fields, $\{\phi_i(l),{\bold m_i(l)}\}$ at every site 
and the argument $l$ denotes imaginary time slice label. The product over $l$ 
from M to 1 implies time ordered products over time slices, with the earlier 
times 
appearing to the right. Finally, the $d^3m_i(l)$ in the integral, implies integration 
over the amplitude and orientation of vector auxiliary fields, ${\bold 
m_i(l)}$.Dropping the $\tau$ dependence of (Aux. F.) allows us to extract an 
effective Hamiltonian from $Z$. To make a further simplification to treat the 
$\phi_i$ Aux. field at its saddle point. While this is not necessary, it reduces the 
number of Aux. fields to be handled per site and leads to efficient computation. 
Thus in the effective Hamiltonian ($H_{eff}$) the fermions couple to the `static' 
HS 
field ${\bold m_i}$ and to the average local charge density. With the redefinition 
${\bold m}_i\rightarrow\frac{U}{2}{\bold m}_i$ we can finally write the effective 
Hamiltonian as:

\begin{eqnarray} 
H_{eff}&=&
H_o+\frac{U}{2} \sum_i(\langle n_i\rangle n_i-{\bold m_i}.{\bold \sigma_i})
\label{ham}\\
&+&\frac{U}{4}\sum_i({\bold m_i}^2-\langle n_i \rangle^2)-\mu\sum_i n_i
\nonumber
\end{eqnarray}
\subsection{Solution of $H_{eff}$ }
$H_{eff}$ coincides with the mean-field Hamiltonian at $T=0$, where ${\bold 
m_i}$ has the interpretation of the local magnetization. However at finite 
temperature these (Aux. F.) do not play the role of magnetization and should 
be 
thought simply as some classical variables (as we have dropped the 
$\tau$ dependence) which can take arbitrary amplitude and angular 
fluctuations.

We simulate $H_{eff}$ by sampling the $\{\bold m_i\}$ fields within a classical 
Monte Carlo (MC) coupled with exact diagonalization for the fermion sector. 
We start the calculation at a high temperature with a random 
configuration 
of $\{\bold m_i\}$ (Aux. F.)'s and uniform on site densities ($\{\langle 
n_i\rangle\}$). For a fixed $\{\bold m_i\}$  
configuration, the Hamiltonian Eq. \ref{ham} is diagonalized. Eigenvectors are 
used to recompute the new $\{\langle n_i\rangle\}$. This process is repeated till 
the self consistent set of $\{\langle n_i\rangle\}$ are obtained. The $\{\langle 
n_i\rangle\}$and the $\{\bold m_i\}$ 
are 
used to compute the free energy of the system. Then as in usual single site 
update scheme the $\{\bold m_i\}$ (Aux. F.) at some site is changed and the 
above 
process is repeated to compute the free energy of the system with the updated 
configuration. Finally a Metropolis algorithm is used to accept/reject the move. 
The goal of our calculation is to generate large number of equilibrium 
configurations of the (Aux. F.) $\{\bold m_i\}$ at a given temperature. These are 
stored so that at any time the eigenvectors/eigenvalues of the full system can 
be 
readily computed without having to rerun the full simulation. The desired density 
of half filling is maintained by adjusting the chemical potential ($\mu$). 

For accessing large system sizes we employ the traveling cluster 
approximation\cite{tca-1,tca-2} 
(TCA) with a 8$^2$ cluster used to anneal at 32$^2$ system. 
All parameters are in units of the hopping $t$.  We employ 4000 MC system 
sweeps among which 2000 are used to thermalize the system, and the rest  for 
calculating observables. We define a MC system sweep to consist sequentially 
visiting every  lattice site  and updating the local ${\bold m_i}$ 
followed by the above mention Metropolis algorithm. 
The local density $\langle n_i \rangle$ is computed from the 
eigenvectors after each diagonalization. We start the calculation at high 
temperature and then gradually cool down to lower temperatures.

We study the formation of local moments as explained below, we start the MC
at $T/t=100$ and cool down in steps of $\Delta{T/t}=10$ up to 10. 
From 
$T/t$=10 to 1,
we use a step size of 1.0. Again the temperature is lowered from 1.0$t$ to 
0.3$t$ 
by grid width 0.1$t$. After that $T/t$ is decreased from 0.3 to 0.1 with interval 
0.05. Then it is made down to 0.01$t$ from 0.1$t$ with spacing of 0.01$t$.
Below this temperature, specifically from 0.01$t$ to 0.005$t$, we 
reduce further
with the interval 0.001$t$. This slow process allows us to avoid getting stuck 
metastable states.
\begin{figure}[t]
	\centering{
		\includegraphics[width=8.5cm, height=3.5cm, clip=true]{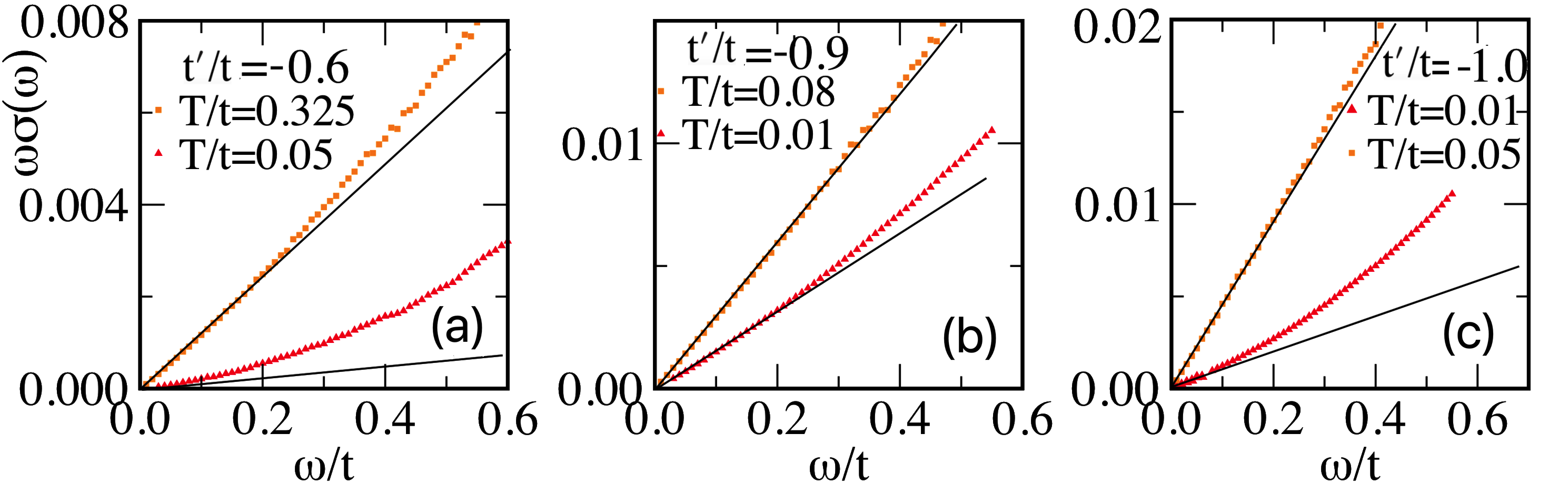}}
	\caption{The  panels  (a) to (c) show the frequency times optical conductivity $\omega\sigma(\omega)$ for $U/t=5.6$  and three $t^{\prime}$ values corresponding to the panel Fig.~\ref{f-2} (b). The $t^{\prime}$ values are mentioned in the panels. In each of the  panels, data is shown for two temperature values. These are discussed in the main text. In these panels solid lines are a guide to eye.}
	\vspace{-0.0cm}
	\label{f-5}
\end{figure}

\subsection{Definitions of the indicators}

We use the static magnetic structure factor ($S_q$), densities of states (DOS), distribution of quantum local moment
and optical conductivity
$\sigma(\omega)$ 
in our study. These are defined as follows. The  DOS is defined as $ N(\omega) 
= 
\sum_{m} \delta(\omega-\omega_m)$, where $\omega_m$ are the eigenvalues 
of 
the fermionic sector and the summation runs up to the total number of 
eigenvalues. $N(\omega)$ is calculated by employing standard Lorentzian 
representation of $\delta$ function. The broadening used for the Lorentzian 
is 
$\sim BW/2N^2$, where $BW$ is the fermionic bandwidth at $U=0$. 200 
$N(\omega)$ samples are obtained from the 2000 system sweeps at every 
temperature. We discard 10 MC steps between measurements to avoid 
self-correlations in the data. The 200 $N(\omega)$ samples are used to obtain 
thermally averaged $\langle N(\omega)\rangle_T$ at a given temperature. These 
are further averaged over data obtained from 10-20 independent runs with 
different random number seeds. Similar process is used for computing 
averages 
of all other observables. The static magnetic structure factor is defined as
\begin{equation}
S_q = \frac{1}{N^2} \displaystyle\sum\limits_{i,j} e^{i {\bold q} \cdot 
({\bold 
r}_{i}-{\bold r}_{j} )} 
\langle {{\bold S_i} \cdot {\bold S_j}}\rangle, 
\end{equation}

The local moment at a site $i$, 
is given by $M_i=\langle 
n_{i\uparrow}-n_{i\downarrow}\rangle =\langle 
n_i\rangle -2\langle n_{i\uparrow}n_{i\downarrow}\rangle=\langle 
\mathbf{S}_{iz}^2 
\rangle=\langle (\mathbf{S}_i.\mathbf{\Omega})^2 \rangle$. Here 
$\langle n_{i\uparrow}+n_{i\downarrow}\rangle =\langle 
n_i\rangle$. For uncorrelated case at half filling $\langle 
n_i\rangle=1$, and $\langle 
n_{i\uparrow}n_{i\downarrow}\rangle\rightarrow\langle 
n_{i\uparrow}\rangle\langle n_{i\downarrow}\rangle$. Further, $\langle 
n_{i,\uparrow}\rangle=\langle n_{i,\downarrow}\rangle=1/2$, implying 
$M=0.5$ for the uncorrelated case. $P(M)=\sum_{M_i}\delta(M-M_i)$ is the 
moment distribution. The distribution $P(M)$ is calculated using 200 configuration of (Auxi.F)s  for a particular temperature.

The real space distribution of magnitudes of the auxiliary field 
$\mathbf{m_i}$ is defined as  $P(|m|)=\sum_i\delta(|m|-|\mathbf{m_i}|)$, where 
$i$ runs over the lattice sites. Similar to the other quantities, the average 
$P(|m|)$ is obtained by averaging over 100 to 200 configurations at a given 
temperature.

The d.c conductivity $\sigma_{dc}$ is estimated by the Kubo-Greenwood
expression \cite{mahan} for the optical conductivity. In a one-electron model
system:
\begin{equation}
\sigma(\omega)=\frac{\pi e^2}{N\hbar a_0}
\sum_{\alpha,\beta} (n_{\alpha} - n_{\beta})
\frac{|f_{\alpha \beta}|^2}{ \epsilon_{\beta} - \epsilon_{\alpha} }
\delta(\omega - (\epsilon_{\beta} - \epsilon_{\alpha})).
\end{equation}
The $f_{\alpha\beta}$ are the matrix elements of the current operator, e.g.,
$\langle \psi_{\alpha} | j_x | \psi_{\beta} \rangle$, and the current operator itself 
(in the tight-binding model) is given by $j_x = i  a_0  \sum_{i, \sigma} 
[t(c^{\dagger}_{{i },\sigma} c^{\phantom{\dagger}}_{i+ a_0\hat{x}, \sigma} - h.c)+
t'(c^{\dagger}_{{i },\sigma} c^{\phantom{\dagger}}_{i+ a_0\hat{x}+a_0\hat{y}, 
\sigma} - h.c)]$. 
The $\psi_{\alpha}$ are single-particle eigenstates, and $\epsilon_{\alpha}$ are 
the corresponding eigenvalues. The $n_{\alpha}=f(\mu - \epsilon_{\alpha})$ are 
Fermi factors. We can compute the low-frequency average, $\sigma_{av}(\mu, 
\Delta \omega, 
N) 
= (\Delta \omega)^{-1} \int_0^{\Delta \omega} 
\sigma(\mu, \omega, N) d\omega$, using periodic boundary conditions in all 
directions. The averaging interval 
is reduced with increasing $N$, with $\Delta \omega \sim B/N$. Here the 
constant 
$B$ is fixed by setting 
$\Delta \omega = 0.008t$ at $N=32^2$. Ideally, the d.c. conductivity is finally 
obtained 
as $\sigma_{dc}(\mu) = {\lim}_{L \rightarrow \infty} { \sigma}_{av}(\mu, B/L, L)$. 
However, 
given the extensive numerical cost of our calculation, we simply use the result of 
$32^2$ system 
as our $\sigma_{dc}(\mu)$. The chemical potential is set to target the required 
electron density $n$. 

\subsection{Metal insulator transition at finite temperature}

In the Fig. 5 from panel (a) to (c) show $\omega\sigma(\omega)$ vs 
$\omega$ for $U/t=5.6$ at three different 
$t^{\prime}/t$ values -0.6, -0.9 and -1 respectively. In each of these 
panels, $\sigma(\omega)$ is shown for two temperature values, one below and one above the metal insulator transition temperature. In (a) and (c), we see non linear dependence of $\omega\sigma(\omega)$ on $\omega$ for the low $T$ cases, 
implying an insulating state. In both these cases, at high $T$ 
there is a clear linear dependence of  $\omega\sigma(\omega)$ on $\omega$ 
signifying a insulator to metal transition with temperature. In panel (b), at 
$t^{\prime}/t=0.9$, $\omega\sigma(\omega)\sim\omega$, for both low and high $T$. By performing extensive numerical calculation for optical conductivity, the finite temperature $T-t'$ phase diagrams in Fig. 2 from panel (a) to (c), shown in the main text, are extracted.

\bibliographystyle{apsrev4-1}
\bibliography{bibliography} 

\begin{thebibliography}{49}
\bibitem [{ran()}]{range}%
  \BibitemOpen
  \href@noop {} {}\bibinfo {note} {The fitting is done for the resistivity data
  starting at $T/t\sim 0.04$, to stay above the $T_N$, i.e., in the thermally
  dominated regime.}\BibitemShut {Stop}%
\end{thebibliography}
\end{document}